\documentclass[%
 preprint,
 amsmath,amssymb,
aps,
pre,
]{revtex4-1}

\usepackage{graphicx}
\usepackage{dcolumn}
\usepackage{bm}
\usepackage{color,ulem} 

\begin{document}

\title{Evidence of spatial embedding in the IPv4 router-level Internet network}

\author{Joshua Parker$^{1}$, Arnold Boedihardjo$^{2}$}
\affiliation{$^1$ University of Maryland, Department of Physics, College Park, Maryland 20742, USA}
\affiliation{$^2$ Geospatial Reasearch Lab, Army Geospatial Center, Alexandria, Virginia 22315, USA}

\date{\today}

\begin{abstract}
Much interest has been taken in understanding the global routing structure of the Internet, both to model and protect the current structures and to modify the structure to improve resilience. These studies rely on trace-routes and algorithmic inference to resolve individual IP addresses into connected routers, yielding a network of routers. Using WHOIS registries, parsing of DNS registries, as well as simple latency-based triangulation, these routers can often be geolocated to at least their country of origin, if not specific regions. In this work, we use node subgraph summary statistics to present evidence that the router-level (IPv4) network is spatially embedded, with the similarity (or dissimilarity) of a node from it's neighbor strongly correlating with the attributes of other routers residing in the same country or region. We discuss these results in context of the recently proposed gravity models of the Internet, as well as the potential application to geolocation inferrence. 
\end{abstract}

\maketitle


\section{Introduction}
Due to the inherent decentralized structure of the global Internet, efforts to generate a comprehensive map of the Internet's connectivity increasingly rely on brute force passive IP address targeting. To be specific, a computer attempts to query (or ``ping") a particular IP address multiple times, each time increasing the time-to-live counter (TTL), a number specifying the maximum unique routers to traverse before terminating. By performing this sequential procedure, the trajectory that would be taken between two routers is recorded. This is called a ``traceroute." By performing millions of traceroutes from different monitoring stations, the various connections between the routers visited can be deduced, and algorithms can be employed to resolve routers with multiple IP addresses~\cite{2010keys}, in the end providing a logical connectivity map of the Internet at the router level. 

In addition to this map, a spatial map of the Internet can be achieved by using a variety of techniques to geolocate a router. Latency triangulation using multiple monitors can restrict the domain to a particular country or set of countries~\cite{2002Eugene,2006gueye,2003tang,2004dabek}. Parsing of DNS registries or entries in the WHOIS database often times yield hints to the country, region, or city specific references. Combining all of these tools with traceroutes has also been seen to be useful for geolocation inference~\cite{2003connolly}. Though these tools are not always accurate, a course map can still be obtained for a large portion of the Internet at the router level.

Currently, the focus of studying the growth of the Internet network focuses on the logical structure at the autonomous systems (AS) level, where nodes are grouped together to represent inter-domain peering relationships between service providers. These studies employ models based on ``preferential attachment" and have been shown to reproduce certain measures of the network, i.e. degree distributions, shortest path, etc~\cite{1999faloutsos}. These models have been used to test efficient routing strategies~\cite{2003krioukov}, as well as to provide insight into the structural integrity of the network when exposed to random failure or intentional attack~\cite{2000albert}. 

However, the assumption behind preferential attachment models have been called into question with regard to Internet connectivity, and traffic-like gravity models for the AS-level Internet have been proposed~\cite{2007bar,2009willinger}. This echoes Tobler's first law of geography, ``Everything is related to everything else, but near things are more related than distant things"~\cite{1970tobler}, implying that local factors in each country may affect the growth of the Internet in that region, meaning that the logical network of the Internet is in fact spatially embedded. This would imply further that the router-level Internet network is also spatially embedded.

In this work, we present evidence of this spatial embedding. Using a subset of the global IPv4 router network, we use node summary characteristics as the node feature space and find that sets of nodes geolocated to a particular country show statistically significant deviation in their mean inter-node distances in this feature space when compared to random sets of nodes. This deviation is bimodal, with most countries showing strong similarity among nodes, while a small subset of countries show strong dissimilarity. This is also true of node regions, but to a lesser extent. We further discuss the implications of our findings on further refinement of geolocation inference.

\section{Data and Analysis}
\subsection{Data sources}
To build our version of the IPv4 Internet network, we relied on the data sources provided by the Center for Applied Internet Data Analysis (CAIDA), namely their Internet Topology Data Kit (ITDK)~\cite{2001claffy}\footnote{The CAIDA UCSD Internet Topology Data Kit - 2014-04,
http://www.caida.org/data/Internet-topology-data-kit}. This data represents connections at the IP address level that were acquired by traceroutes from around the world, and assigned router identities based on various IP alias techniques. This data comes in ``links", i.e. lists of routers and router interfaces seen to be linked in the data set. Therefore, we considered the presence of two routers in a link to be connected without direction, yielding a binary, undirected network. Along with this connectivity information, CAIDA's ITDK supplies geolocation acquired from the MaxMind Geolite database for a subset of the routers to varying levels of precision. The error in the MaxMind database is negligible (1 percent) when considering only the country, and country specific in it's reliability when interested in a node's region. Table 1 summarizes some basic properties of this network as well as the amount of geolocation information provided. 

\begin{table}[!h]
\centering
    \caption{Basic properties of the network analyzed}
    \begin{tabular}{c r c c r}
	\multicolumn{2}{c}{\textbf{Network Statistics}}& & \multicolumn{2}{c}{\textbf{Level of node geolocation}}\\    
\hline
	\\
    Nodes & 3,248,358 & & None & 1,520,465 \\
    Edges & 14,083,946 & & Only country & 604,939 \\
              &            & & Country and region & 1,122,954 \\
    \end{tabular}
\end{table}

\subsection{Node feature space}
For each node, we developed a feature set out of node summary statistics based on the local network neighborhood. The first variable was simply the degree of a node $k_i = |\{e_{ij}\}|$, i.e. the cardinality of the set of all edges incident to node $n_i$. Next, considering the set comprised of the nodes $n_j$ that are in the neighborhood of node $n_i$, we calculate the average neighbor degree ($\langle k_j \rangle_i = \frac{1}{k_i}\Sigma_j k_j$), the local neighbor degree variance with respect to the global average ($\sigma_i = \frac{1}{k_i -1} \Sigma_j (k_j - \langle k \rangle)^2$), and the local neighbor degree correlation with respect to the global average and global standard deviation ($p_i = \frac{1}{\sigma_k^2 k_i} \Sigma_j\Sigma_i(k_i - \langle k \rangle)(k_j-\langle k \rangle)$). These values were computed for all nodes, regardless of their geolocation status, using software written in Java and C/C++. 

\subsection{Analysis and Statistical Tests}
\begin{figure}[ht]
\includegraphics[width=.5\textwidth]{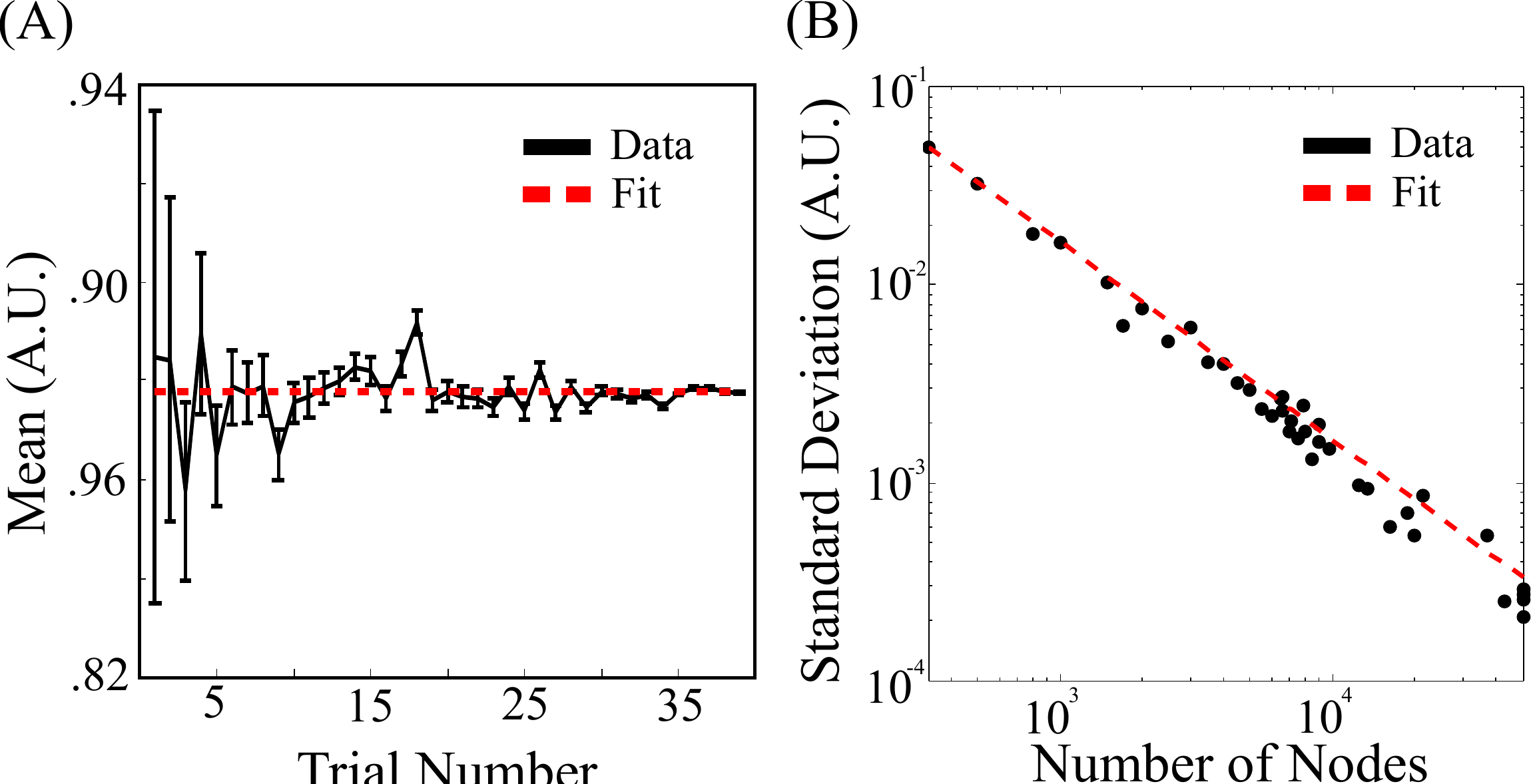}
\caption{\label{Fig1} (Color online) Results of random samplings of 40 sets of variable set size, both the mean (left) and standard deviation (right) of inter-node distance calculations. Red dashed line represents fit}
\end{figure}

Since these variables are not necessarily orthogonal, we computed the principal components of the data set and used a node's transformed variables (using scores on the four principle components), allowing us to calculate the Mahalanobis distance between nodes by applying a Euclidean norm~\cite{1936mahalanobis}. To develop our statistical tests, we generated 100 random node sets of size N, calculated inter-node distances, and took an average and standard deviation. Doing this for sequentially larger subsets of nodes showed the predictable convergence, consistent with the central limit theorem applied to a distance distribution with stable mean $\mu_r = .877654$ and standard deviation that scales as $\sigma_r(N) = 16.45 N^{-1}$ (see Fig.\ref{Fig1}). Using this information, we then measure the mean inter-node distance for nodes geolocated in the same country (using all the country's nodes) and calculate the Z-score, $z = (\mu_{data} - \mu_r)/\sigma_r(N_{data})$. This allows us to discuss the statistical significance of the inter-node distances in each country by testing it against the null hypothesis of each country's node set being essential random. 

\section{Results and Discussion}
\begin{figure}[ht]
\includegraphics[width=.5\textwidth]{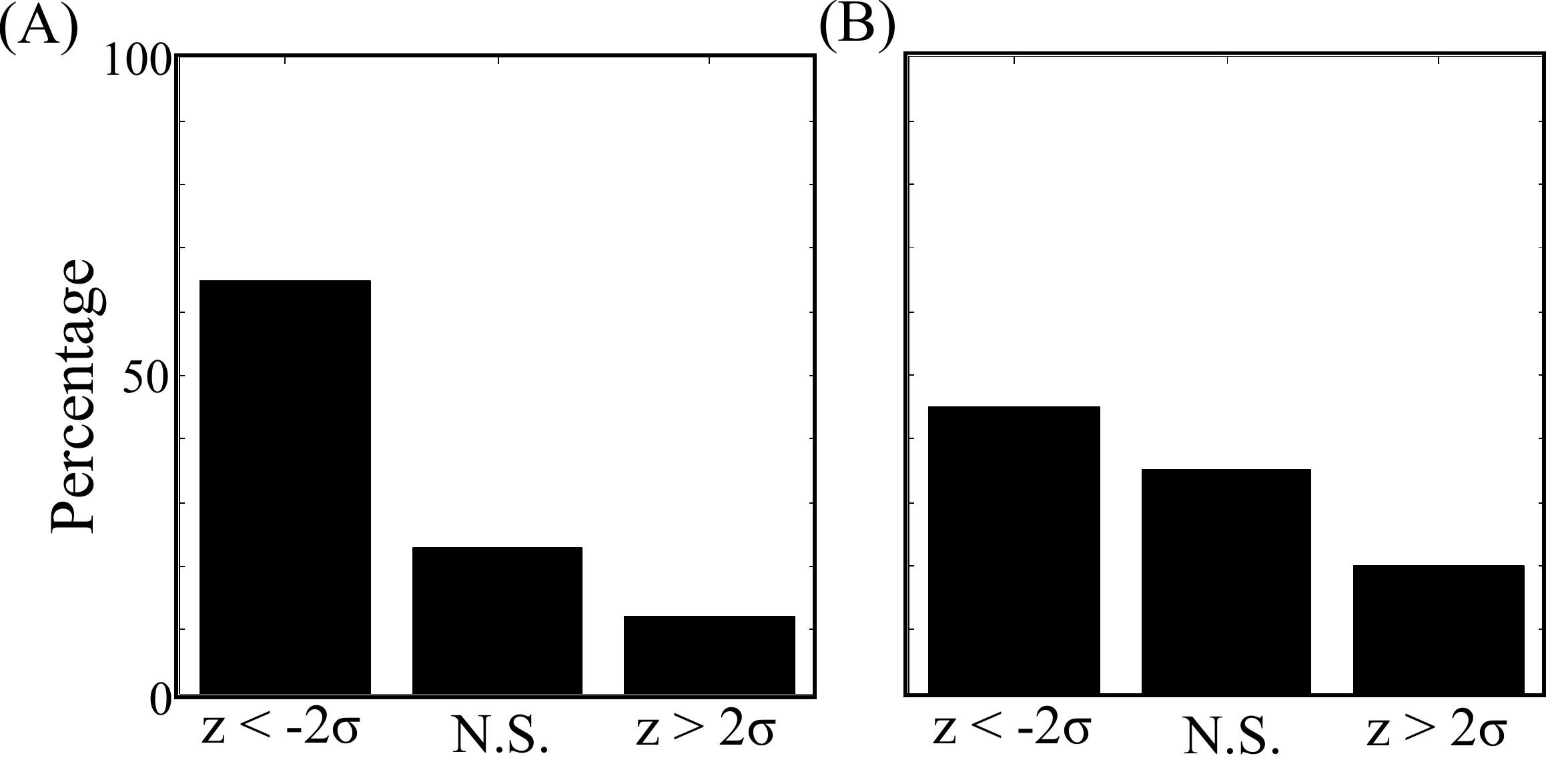}
\caption{\label{Fig2} Results of statistical tests on the countries (left) and regions (right). $|z| > 2\sigma$ implies statistical significance at the $ p < .05$ level}
\end{figure}

Fig.\ref{Fig2}.A shows the distribution Z-scores for the 180 countries represented in the data set. As can be seen, the large majority of countries show deviation from the null hypothesis ($p < 0.05$), suggesting the presence of spatial embedding. Furthermore, this deviation is bimodal in that some countries show strong inter-node similarity ($z << 0$) while others show strong dissimilarity ($z >> 0$). This is also true, albeit to a lesser extent, when considering nodes within the same region (354 regions in all, see Fig.\ref{Fig2}.B). In regards to the Internet, Tobler's first law seems to hold.
\subsection{Analysis and Statistical Tests}

These results suggest that the logical connectivity map of the Internet is spatially embedded, with the correlations or anti-correlations between variables geolocated to the same place deviating sharply from randomly selected nodes. This supports the notion that a gravity model for Internet growth is more appropriate than random attachment, where model parameters and growth behavior should include strong geographic elements. To this end, further research needs to be done to elucidate what geographic, social, and economic forces drive the heterogeneous growth of the Internet across countries and regions so that their effects can be quantitated in future Internet growth models. This suggests a strong role for geographers in Internet cartography research.

Currently, we are investigating to what extent the similarity or dissimilarity of a node from other nodes can be used to infer a geolocation for un-located nodes. If countries always had ``close" nodes in terms of our defined metric space, then simply minimizing a nodes average distance to the nodes of a particular country would suffice. The bi-modality of the distribution, however, suggests that enforcing conditions on the variance of the inter-node distance distributions would be more appropriate, but this work is still in progress. We are also investigating whether additional, more complex node summary statistics such as the local clustering coefficient would further distinguish countries and regions from random sets.

\section*{Acknowledgment}

The authors would like to thank Alexander Yale-Loehr, Sean Sovine, Harland Yu, Crystal Chen, Raimundo Dos Santos, and Nicole Wayant for their questions and comments during the process of this research. JP would like to thank the ASEE for financial support.

\bibliographystyle{IEEEtran}
\bibliography{References}
 \end{document}